\documentclass[prb,aps,floats,amssymb,showkeys,showpacs,
superscriptaddress]{revtex4}
\usepackage{graphicx}
\usepackage{graphics}
\usepackage{amsmath}
\usepackage{epsfig,bm}
\usepackage{rotating}
\pdfoutput=1
\begin{document}

\title{A positivity property of the dimer entropy of  graphs}

\author{P. Butera}
\email{paolo.butera@mib.infn.it}
\affiliation
{Dipartimento di Fisica Universita' di Milano-Bicocca\\
and\\
Istituto Nazionale di Fisica Nucleare \\
Sezione di Milano-Bicocca\\
 3 Piazza della Scienza, 20126 Milano, Italy}

\author{P. Federbush}
\email{pfed@umich.edu}
\affiliation
{Department of Mathematics\\
University of Michigan \\
Ann Arbor, MI 48109-1043, USA\\}

\author{M. Pernici} 
\email{mario.pernici@mi.infn.it}
\affiliation
{Istituto Nazionale di Fisica Nucleare \\
Sezione di Milano\\
 16 Via Celoria, 20133 Milano, Italy}

\date{\today}
\begin{abstract}

  The entropy of a monomer-dimer system on an infinite bipartite
  lattice can be written as a mean-field part plus a series expansion
  in the dimer density. In a previous paper it has been conjectured
  that all coefficients of this series are positive.  Analogously on a
  connected regular graph with $v$ vertices, the ``entropy'' of the
  graph ${\rm ln} N(i)/v$, where $N(i)$ is the number of ways of
  setting down $i$ dimers on the graph, can be written as a part
  depending only on the number of the dimer configurations over the
  completed graph plus a Newton series in the dimer density on the
  graph.  In this paper, we investigate for which connected regular
  graphs all the coefficients of the Newton series are positive (for
  short, these graphs will be called positive).  In the class of
  connected regular bipartite graphs, up to $v=20$, the only non
  positive graphs have vertices of degree $3$.  From $v=14$ to $v=30$,
  the frequency of the positivity violations in the $3$-regular graphs
  decreases with increasing $v$.  In the case of connected $4$-regular
  bipartite graphs, the first violations occur in two out of the
  $2806490$ graphs with $v=22$.  We conjecture that for each degree
  $r$ the frequency of the violations, in the class of the $r-$regular
  bipartite graphs, goes to zero as $v$ tends to infinity.
  
  This graph-positivity property can be extended to non-regular or
  non-bipartite graphs.

  We have examined a large number of rectangular grids of size $N_x
 \times N_y $ both with open and periodic boundary conditions.  We
 have observed positivity violations only for $min(N_x, N_y) = 3$ or
 $4$.

\end{abstract}
\maketitle

\section{ Introduction} 
For periodic cubical $d$-dimensional lattices with even length
sides, the number of dimer configurations covering a fraction $p$ of
the vertices (called ``matchings'' of the lattice graph in the
language of graph theory) has long been studied. As the number of
vertices $v$ tends to infinity, the number of distinct dimer
configurations grows[\onlinecite{hamm}] as $\exp(v\lambda_d(p))$, where
$\lambda_d(p)$ is called the dimer entropy and $p$ is the dimer density.
In [\onlinecite{FF}],
the entropy has been written as the sum of a mean field term and of a
series expansion in the dimer density, computed through order $6$; in
Refs.[\onlinecite{bpyl,bfp}] this series has been extended at least
through the order $20$. 
More precisely, for lattices of dimension $2
\le d \le 4$ the first $24$ series coefficients were derived[
\onlinecite{bpyl,bfp}],
while 20 coefficients have been obtained for all $ d> 4$
(in $d=5, 6$ respectively $22, 21$ coefficients have been computed).
In the case of
$d=1$, all coefficients are positive and have long been known.  It is
striking that, for any integer $d$, all the series coefficients that
we could compute, are positive. This led us to conjecture[\onlinecite{bfp}]
that: {\it for all (hyper)-cubical lattices the series coefficients
are all positive.} Moreover, after examining the series coefficients
for various non-square bipartite lattices, we were led to extend the
conjecture to all bipartite lattices (i.e. to the lattices in which
the sites can be divided into even and odd, such that no edge connects
two vertices with the same parity), while the positivity does not
hold in the case of non-bipartite lattices.

We observe that, in analogy with the case of infinite bipartite
lattices, on a connected simple undirected regular bipartite graph $G$
with set of vertices $V$ and set of edges $E$, the ``graph dimer
entropy'' ${\rm ln} N(i)/v$ - where $N(i)$ is the number of ways of
setting down $i$ dimers on the graph and $v=|V|$ is the number of
vertices of the graph - can be written as a part depending on the
dimer configurations counting over the completion of the graph and a
Newton expansion in the dimer density on the graph. If all the
coefficients of the Newton series are positive, we shall say that the
graph satisfies the {\it graph-positivity} property.  For not too
large number of vertices $v$, we have generated {\it all} regular
(bi)connected bipartite graphs (RBB graphs from now on) using the {\it
Nauty} program[\onlinecite{nauty}] via the [{\it Sage}] 
interface[\onlinecite{Sage}].
In examining all RBB graphs with vertices of degree $3$ for $v \le
30$, we have observed no violations of the graph positivity for $v <
14$ vertices.  A single violation is observed for $v=14$. For $v \ge
14$, as the number $v$ of vertices increases, the frequency of
violations decreases.  No positivity violation occurs in
all RBB graphs with vertices of degree greater than $3$, up to $v=20$.  
For $v=22$ and degree $4$ there are two violations.  For
$r-$regular graphs with $r>4$ the absence of violations could be checked up
to $v=20$.  Although we found no positivity violation for RBB graphs
with vertex degree larger than $4$ in the large sample studied, we can point
out examples of positivity violation occurring for larger values of
$v$, in some $r-$regular graphs with $r=5,6,7,8$, which were defined
in [\onlinecite{Wanless}].  We conjecture that, for RBB graphs
of any degree {\it the
frequency of positivity violations becomes vanishingly small as $v$
becomes large}.

An interesting observation emerging from the systematic study of RBB
graphs with $v \le 30$ and degree $3$, is that the average order of
the automorphism group of the graphs not satisfying the
graph-positivity property is larger by an order of magnitude than the average 
order of the automorphism group for all RBB graphs.

In addition to this systematic survey of RBB graphs, we have focused
the attention on the class of square and hexagonal lattices with
periodic boundary conditions (bc), which are also RBB graphs.  In the case
of rectangular grids of sizes $N_x \times N_y$ with periodic bc,
 we examined the cases $N_x=N_y$ up to sizes $N_y=12$. For
$N_x \ge N_y$ we found positivity violations only for $N_y=4$ and $N_x
\ge 432$.

For periodic hexagonal lattices of
sizes $N_x \times N_y$ in the brick-wall representation, positivity is valid
when $N_x \ge N_y$, apart from the case $N_x=N_y=4$,
while violations occur for $N_x < N_y$.  We have tested
this class of grids up to the size $14 \times 14$.

This positivity property can also be studied for graphs which are not RBB.
Since we originally proposed the positivity conjecture for infinite
square lattices, it is natural to consider also rectangular grids with
open b. c..  
In the case of rectangular grids $N_x \times N_y$ with open bc,
in the cases $N_x = N_y$ we considered up to the case $N_y=19$;
the graph positivity holds except for the case of $N_x \times 3$ graphs.
We have also examined the case of rectangular grids $N_x \times N_y$
with bc periodic only in the $y$ direction. Again, we
found positivity violations only for narrow grids, as in the case of
rectangular grids with periodic and with open bc.

We considered also several cubic grids of sizes $N_x\times N_y\times N_z$
with open bc, up to the $5\times 5\times 4$ case: 
 no positivity violation was observed, except in the $N_x\times 2\times 2$
 case, which is isomorphic to the $N_y=4$ rectangular grid case periodic 
 in the $y$-direction.

In the case of non-bipartite graphs, the positivity property is not
common. However there are exceptions for specific classes of graphs:
for example, we have singled out a sequence of ``nanotubes'' $C_{40+20
N}$, with $N$ the number of hexagonal strips, and have tested graph
positivity for $1 \le N \le 300$.  All the nanotubes with $N > 7$ are
graph-positive.

To perform our study of graphs, it was necessary to compute the graph
matching generating polynomial
\begin{equation}
M(t) = \sum_{i=0}^{[v/2]} N(i)t^i
\end{equation}
where $N(i)$ is the number of configurations of $i$ dimers on the graph.
We used the algorithm discussed in [{\onlinecite{bpalg}] to perform the 
computation.

The paper is organized as follows. In the second Section, we formulate
the graph positivity property.  The third Section summarizes the
results of the graph tests for a variety of graphs and lattices. The
fourth Section contains our conclusions. In Appendix A, the graph
positivity is proven for polygons, for complete bipartite graphs and
for an approximate average distribution of graphs.  In Appendix B the
positivity property is examined for a sequence of nanotubes.
In Appendix C we give a few examples of RBB graphs violating positivity.

\section{The positivity property for graphs}

To formulate the graph-positivity properly, let us turn now to a cubic
lattice graph $G$ with $v$ vertices of degree $2d$ and periodic
bc. Let $N(i)$ be
the number of ways of setting down $i$ dimers on the edges of $G$
(with no overlap).  If we consider the sequence of larger and larger
periodic lattices used to compute the entropy $\lambda_d(p)$ on the
infinite lattice, one must have

\begin{equation}
    \hat{\lambda_d}(\hat{p_i}) = \frac {1}{v} {\rm ln}N(i) \to  
\lim_{v \to \infty} \frac {1}{v} {\rm ln}(\exp v\lambda_d)
=\lambda_d(p)
\label{ln}
\end{equation}
 where the arrow indicates convergence as $v \to \infty$.
Here $i$ is related to $p$ by 

\begin{equation}
    \hat{p_i} = \frac{2i}{v} \approx p
\label{ip}
\end{equation}
where the integer $i$ is chosen to make this approximation best. We refer now
 to the notation in [\onlinecite{FF}] observing that from Eq.(5.9) of this 
\begin{equation}
\lambda_d(p)=\frac{p}{2} {\rm ln}(2d) +\lim_{v \to \infty}  \frac {1}{v}{\rm ln}Z 
\label{ip2}
\end{equation}

Observe now that $Z$ can be factored into a ``mean field'' term $Z_0$ 
times the rest  $Z^*$ as in Eq.(5.12) of  [\onlinecite{FF}] 
\begin{equation}
  \lambda_d(p)=\frac{p}{2} {\rm ln}(2d) 
 +\lim_{v\to \infty}  \frac {1}{v}{\rm ln}(Z_0)
  +\lim_{v \to \infty}  \frac {1}{v}{\rm ln}(Z^*) . 
\label{ip3}
\end{equation}

\noindent
  From Eq.(\ref{ln}) and from Eq.(5.11) 
and (7.1) of [\onlinecite{FF}], we have 
\begin{equation}
    \frac {1}{v} {\rm ln}\frac{N(i)}{(2d)^i} - \lim_{v \to \infty} \frac
  {1}{v}{\rm ln}(Z_0) \to \sum_{k=2}a_k(d)p^k
\label{ip6}
\end{equation}
\noindent
 with
\begin{equation}
  \lim_{v \to \infty} \frac {1}{v}{\rm ln}(Z_0)= -\frac{p}{2} {\rm
  ln}(p)- (1-p){\rm ln}(1-p) - \frac{p}{2}.
\label{ip7}
\end{equation}

It is an easy computation to show that for the graph $\bar G$, defined
as the completion of $G$ (namely the non-bipartite graph constructed
by connecting the vertices of $G$ in all possible ways), if we set
$\bar N(i)$ as the number of ways of setting down $i$ dimers on $\bar
G$ (without overlap), 

\begin{equation}
    \bar N(i) = \frac{v!}{(v-2i)!i!2^i}
\label{nbar}
\end{equation}

then
\begin{equation}
\frac {1}{v} {\rm ln}\frac{\bar N(i)}{(v-1)^i}  \to \lim_{v \to \infty}  \frac {1}{v}{\rm ln}(Z_0)
\label{ip8}
\end{equation}
\noindent
We now define

\begin{equation}
 d(i)
 \equiv  {\rm ln}\frac{ N(i)}{(2d)^i} - {\rm ln}\frac{\bar N(i)}{(v-1)^i} 
\label{ip9}
\end{equation}
\noindent
and see that 
\begin{equation}
\frac{1}{v} d(i) \to \sum_{k=2}a_k(d)p^k
\label{ip10}
\end{equation}
\noindent
The rhs of Eq.(\ref{ip10}) is 
the series part of the dimer entropy in Eq.(6) in [\onlinecite{bfp}],
where it is conjectured that the $a_k$ are all positive.
We now observe that as $v \to
\infty$, the finite differences at $i=0$ on the lhs of 
Eq. (\ref{ip10}) become the derivatives on the rhs.

 We write $\Delta$
for the forward finite difference with respect to $i$.  One has
\begin{equation}
     \frac{v}{2}\Delta \approx \frac {d}{dp}
\label{ip11}
\end{equation}
\noindent
 so differentiating Eq.(\ref{ip10}) once, one gets
\begin{equation}
\frac{1}{2}( d(i+1)-d(i)) \to \frac {d}{dp} \sum_{k=2}a_k(d)p^k.
\label{ip12}
\end{equation}
\noindent
%Continuing to differentiate we see that all finite difference
%derivatives of $d(i)$ are $\geq 0$.

Writing $d(i) = \hat{d}(\hat{p})$ with $\hat{p} = \frac{2 i}{v}$
and $h=\frac{2}{v}$, 
one has the Newton expansion

\begin{equation}
    \hat{d}(\hat{p}) = \sum_{k=0}^{v/2} \frac{\Delta_h^k[\hat{d}](0)}{k!} (\hat{p})_k
\end{equation}
where 
$(\hat{p})_k = \hat{p}(\hat{p}-h)...(\hat{p}-(k-1)h) = (\frac{2}{v})^k (i)_k$, 
 $(i)_k$ is the falling factorial and $\Delta_h$ is the finite difference
 with step $h$, 
 $\Delta_h \hat{d}(\hat{p}) = \frac{\hat{d}(\hat{p} + h) - \hat{d}(\hat{p})}{h}$.

We can now state the definition of graph positivity in greater
generality.  
Let $G$ be a simple regular biconnected graph, with $v$
vertices of degree $r$. 
Let $\bar G$ be the completion
of $G$. Let $N(i)$ be the number of ways of laying down
$i$ dimers on $G$ (without overlap) and $\bar N(i)$ the same quantity
for $\bar G$. The equations introduced in this section for the hypercubic
lattices generalize to  regular biconnected graphs
replacing $2d$ with the degree $r$.

Define

\begin{equation}
    \hat{d}(\hat{p}) = d(i)
 = {\rm ln}\frac{N(i)}{r^i}  - {\rm ln}\frac{\bar N(i) }{(v-1)^i} .
\label{ip13}
\end{equation}
and
\begin{equation}
    \hat{\lambda}(\hat{p_i}) = R(i) +
    \frac{h}{2} \sum_{k=0}^{v/2} \frac{\Delta_h^k[\hat{d}](0)}{k!} (\hat{p})_k
\label{entropy}
\end{equation}
where
\begin{equation}
    R(i) =\frac{ 1}{v} {\rm ln}\left(\frac{r^i}{(v-1)^i}\bar N(i)\right)
\end{equation}

Using the Stirling formula for large $v$
\begin{equation}
R(i) \approx \frac{1}{2}(p{\rm ln}(r) - p{\rm ln}(p) -2(1-p){\rm ln}(1-p) - p)
\end{equation}

Then $G$ satisfies graph positivity, if all the $d(i)$ are
non-negative, as are all terms derived by any number of non-trivial
finite differences i.e. if all terms in the sequence
$d(0), d(1),d(2),d(3),...$ are non-negative as are all terms in the sequence
$d(1) - d(0), d(2)-d(1), d(3)-d(2),....$, and the sequence 
$d(2)-2d(1)+d(0), d(3)-2d(2)+d(1), d(4)-2d(3)+d(2),...$ etc., that is
\begin{equation}
\Delta^k d(i) \ge 0
\label{dbound}
\end{equation}
for $k=0,...,\nu$ and $i=0,...,\nu - k$.
In fact, it is sufficient that the first term $\Delta^k d(0)$
in each of these sequences is non-negative, non-negativity of the
other terms would then follow ($\Delta^k d(0) = 0$ for $k=0,1$ since
$d(0) = d(1) = 0$).
Equivalently, positivity means that all
the coefficients in the Newton series of $\hat{d}(\hat{p})$ are
non-negative.

In all the tests we made,
when a graph is positive it has $\Delta^k d(i)$ strictly positive
apart from the trivial case $k=0, i=0,1$ and $k=1, i=0$.

For infinite regular lattices, in the bipartite models
$a_k > 0$ while it is not true for non-bipartite ones, so
we can expect that regular connected bipartite (hence ``biconnected'') 
graphs (RBB graphs) tend to be positive, while non-bipartite graphs not.

Notice that using $N(1) = \frac{v r}{2}$ one can rewrite
$d(i)$ in Eq.(\ref{ip13}) with no explicit dependence on degree of the vertices
\begin{equation}
    d(i) = {\rm ln}\frac{N(i)}{N(1)^i}
- {\rm ln}\frac{\bar{N}(i)}{\bar{N}(1)^i}
\label{di}
\end{equation}
\noindent
for $i=1,..,\nu$, where $\nu$ is the maximum value of $i$ with $N(i)$ not
zero, and  $\bar{N}(i)$ is the number of configurations of $i$ dimers on a complete
graph with $\bar{v}=2\nu$ vertices (if $\bar{v}$ = $v$ the graph has
perfect matchings).  This suggests that it is worthwhile to
investigate also when the graph-positivity holds for non-regular graphs.

Let us comment on the use of the graph completion in (\ref{ip8}).
If one studies the definition of $Z_0$ in [\onlinecite{FF}], the
expression on the lhs of (\ref{ip8}) is seen to arise naturally. But
we get no theoretical understanding of why the completion of a graph,
rather than the bipartite completion, is taken. 
One could have
in (\ref{ip8}) used a similar expression with the bipartite
completion, but one can show that there are more violations of the bounds
Eq.(\ref{dbound}) for $k \ge 2$ than with the definition used;
in particular the RBB graphs with $v \le 18$ would all violate
positivity, apart from the complete bipartite graphs, which have
trivially $d(i) = 0$.

\section{Tests of graph positivity} 
As a first step, we have tested systematically the validity of the
positivity property on RBB graphs, exploring them in ascending number
of vertices, namely we have counted the configurations of $i$ dimers
on these graphs and checked the inequalities for the higher finite
differences.  To generate systematically the RBB graphs, we have used
the {\it geng} program in the { \it Nauty} package[\onlinecite{nauty}], via
the {\it Sage} interface[\onlinecite{Sage}]. We have used 
an ordinary desktop personal computer
based on a processor Intel $i7$ $860$ 
and a RAM of 8 $GB$.
The first result of this extensive graph survey is that, while most RBB
graphs share the graph-positivity property, a few of them do not, see
 Table \ref{tabdeg3} for the graphs with degree $3$. 
 There are no violations for $v < 14$.
The  positivity test for graphs with $30$ vertices  was completed in
about $20$ days. Most of the computing time was spent in enumerating 
graphs with   {\it geng}.

\begin{table}[ht]
 \caption{For RBB graphs with $14 \le v \le 30$ vertices of degree 3, 
 we have listed
 the number of graphs in this class, the number of violations of the
 graph positivity, the average order $ng$ of the automorphism group of
 graphs, the average order $ngv$ of this group for graphs
 violating the positivity property.}
\begin{tabular}{|c|c|c|c|c|}
 \hline
 $v$ & number of graphs & violations & $ng$ & $ngv$\\
 \hline
 14&    13&   1 & 44.5 & 336.\\
 16&    38&   2&  18.6 & 112.\\
 18&    149&  2 & 15.1 & 176.\\
 20&    703&  8 & 8.7  & 118.\\
 22&   4132&  17 & 4.5  & 72.\\
 24&  29579&  49 & 3.3 & 92.\\
 26& 245627&  115 & 2.3 & 66.\\
 28& 2291589& 514 & 1.9 & 55.\\
 30& 23466857& 3949 & 1.7 & 37.\\
 \hline
 \end{tabular}
 \label{tabdeg3}
\end{table}

For  $3$-regular graphs, the frequency of violations decreases
with $v$, for $v \ge 14$. 

It is interesting to observe that, in the cases considered in 
Table \ref{tabdeg3}, the average order of the automorphism group of
the positivity-violating graphs is roughly an order of magnitude
greater than the average order of the automorphism group of all the
RBB graphs with the same number of vertices. 
In particular,
the positivity violating graph with $v=14$ is the most symmetrical of all
$v=14$ RBB graphs with degree $3$.

For degrees greater than $3$, there is no violation of positivity for
RBB graphs with number of vertices up to $v=20$.  For $v=22$, there
are $2806490$ RBB graphs with vertices of degree $4$, and two violations,
given in Appendix C.
The average order of the automorphism group of the positivity-violating graphs 
is $1080$, the average order of the automorphism group of all the
RBB graphs with $v=22$ and degree equal to $4$ is $1.5$.

Although in this systematic examination of graphs with low order we
found no positivity-violating graphs with degree greater than $4$, we
can exhibit cases of positivity-violating graphs with degree up to $8$,
belonging to the class of $k$-regular graphs $G_{k, f}$ considered in
Example $1$ of [\onlinecite{Wanless}].  The $G_{k,f}$ graphs, with
$f \ge 2$ and $k \ge 3$ are RBB graphs with degree $k$ and $v=2n$
vertices where $n=(k^2f+1)$.  The vertex classes of $G_{k,f}$ are
\begin{equation}
    U = \{s\} \cup \{u_{a,b,c} : 1 \le a \le k; 1 \le b \le k; 1 \le c \le f\}
\end{equation}
 and
\begin{equation}
V = \{t\} \cup \{v_{a,b,c} : 1 \le a \le k; 1 \le b \le k; 1 \le c \le f\}
\end{equation}
The edges are $E = E_1 \cup E_2 \cup E_3$ with
\begin{eqnarray}
E_1 &=& \{(u_{a,b,c},v_{a,\beta,c}) : 1 \le a,b,\beta \le k; (b,\beta) \neq (1,1);
1 \le c \le f\} \nonumber \\
E_2 &=& \{(u_{a,1,c},v_{a,1,c+1}) : 1 \le a \le k; 1 \le c \le f-1\} 
\nonumber \\
E_3 &=& \{(s,v_{a,1,1}) : 1 \le a \le k\} \cup \{(u_{a,1,f},t) : 1 \le a \le k\}
\nonumber \\
\end{eqnarray}
\noindent
and dimer countings for almost complete and complete matching
\begin{equation}
\begin{split}
N(n-1) &= \frac{((k-1)!)^{kf}}{(k-2)^2}((k+2)^2(k-1)^{kf+2} + \\
 &((k-2)^2kf - 3k^2 + 4)k^2 (k-1)^{(k-1)f} + \\
 &2k^3(k-1)^{(k-2)f+1})
\label{n1wan}
\end{split}
\end{equation}

\begin{equation}
N(n) = k((k-1)!)^{kf}(k-1)^{(k-1)f}
\label{nwan}
\end{equation}

The graph
$G_{5, 2}$ with $102$ vertices of degree $5$ and the graph $G_{6, 2}$
with $146$ vertices of degree $6$ show a first violation in the case
of $\Delta^2 d(i)$ for some $i > 0$. In the finite differences at $i=0$, a first
violation occurs for $\Delta^{13} d(0)$. 
The graphs $G_{7, 2}$ and $G_{8, 2}$ have $\Delta^{12} d(0) < 0$.
The symmetry groups of the graphs $G_{k, 2}$ with $k=5,..,8$ 
are larger than $10^{29}$.

While all the examples of violations $\Delta^j d(i) < 0$ mentioned so
far have $j \ge 2$, there are violations for $j=1$. $\Delta d(i) \ge 0$
is equivalent to
\begin{equation}
\frac{N(i)}{N(i+1)} \le \frac{(2n-1)(i+1)}{r(n-i)(2n-2i-1)}
\label{ineq2}
\end{equation}
where $v=2n$ and $r$ is the degree. 
This implies $N(n-1)/N(n) \le n(2n-1)/r$.  From
Eqs.(\ref{n1wan}, \ref{nwan}) it follows that the graph $G_{3, 6}$
with $110$ vertices fails to satisfy this inequality.

The $0$th order  inequality $d(i) \ge 0$ gives
\begin{equation}
    N(i) \ge \frac{v! r^i}{(v-2i)!2^ii!(v-1)^i}
\label{ineq1}
\end{equation}
where $r$ is the degree of the vertices. This inequality should be compared to
the ``Lower Matching Conjecture''[\onlinecite{fkm}]
\begin{equation}
N(i) \ge \binom{n}{i}^2 \left(\frac{nr-i}{nr}\right)^{nr-i}\left(\frac{ir}{n}\right)^i.
\label{ineq3}
\end{equation}
recently proven in [\onlinecite{csi}].  In fact Csikvari proves a stronger
bound with the right side of (\ref{ineq3}) divided by $\binom{n}{i}(i/n)^i(1-i/n)^{(n-i)}$.
We thank the referee for informing us that he can
prove (\ref{ineq1}) is true provided either $r \le 5$ or $v \ge 2r^2$.

Gurvits has proved the following  slightly weaker lower bound, 
see Eq.(51) of [\onlinecite{G}].
\begin{equation*}
N(i) \ge\frac{((r-i/n)/r)^{nr-i}(1-1/n)^{(1-1/n)(2n^2)(1-i/n)}}
{(i/(nr))^{i}n^{-2n(1-i/n)}((n-i)!)^2}
\end{equation*}

We do not know whether there are violations of the inequality
Eq.(\ref{ineq1}).

Let us now consider a sequence of increasingly larger
finite rectangular lattices with bc periodic in the 
$x$ and $y$ directions, as
are used in the limiting procedure to extract $\lambda_d(p)$ for
$d=2$.  For this class of graphs, we have found no violation of
graph-positivity for grids of sizes  $(N_x \times N_y)$, with $N_x$
even, such that

$4 \le N_x \le 430, N_y=4$

$6 \le N_x \le 800,  N_y=6$

$8 \le N_x \le 100,  N_y=8$

and for the grids of sizes $N_x \times  N_y= 10 \times 10, 12 \times 10,
 12 \times 12$.

However all graphs are not positive for
$430 < N_x \le 5000, N_y=4$.
\noindent

We have examined also the hexagonal lattices of size $N_x \times N_y$
in the brick-wall representation. 
We have examined the following cases with $N_x \le N_y$, with $N_x$ even,

$4 \le N_x \le 800, N_y=4$

$6 \le N_x \le 270, N_y=6$

$8 \le N_x \le 90, N_y=8$

$10 \le N_x \le 14, N_y=10$

and $(N_x,N_y) = (12,12), (14,12), (14,14)$.
The only violation occurs in the case 
 $(N_x,N_y)=(4,4)$, which represents one of the two
non positive 3-regular RBB graphs with $16$ vertices.

For $N_x < N_y$, we have considered the following cases 

$N_x=4, 6 \le N_y \le 100$, in which all the graphs are non positive.

$N_x=6, 8 \le N_y \le 100$, in which all the graphs $N_y \ge 26$ 
are non positive.

The lattice $4 \times 6 $, a graph with $24$ vertices, is one of the $49$
cases of violations observed in the systematic survey of the 3-regular RBB
graphs with $24$ vertices.

Let us now consider what happens in the case of non-RBB graphs, as
mentioned at the end of Section II.  In the case of disconnected
graphs, the matching generating polynomial is the product of the
matching generating polynomials of the connected components. Many
violations of the graph positivity property are observed. For instance, 
the graph formed by $n$ hexagons, with $n \le 100$, violates
positivity for $n \ge 3$.

For non-regular bipartite connected graphs there are many violations;
for $v$ even up to $12$ the frequency of violations increases with $v$;
for $v=12$ more than  one third of the graphs violate positivity.

For non-regular bipartite biconnected graphs there are in general many
violations; for instance enumerating the bipartite biconnected graphs
with minimum vertex degree $2$ and maximum degree $3$, the frequency of
violations increases with $v$ for $v=10,...,18$.

These violations are due mostly to the vertices of degree $2$.
We examined a couple of classes of cases excluding vertices of degree $2$:
bipartite biconnected graphs with minimum degree $3$ and maximum
degree $4$ have a similar pattern as RBB graphs, see Table
\ref{tabdeg34}: the frequency of violations decreases with $v$ and, on
the average, the non-positive graphs have significantly larger
symmetries;
bipartite biconnected graphs with minimum degree $4$ and maximum
degree $5$ have no violations for $v < 16$; there is one violation
out of $189853$ graphs for $v=16$, no violations for $v=17$ out of
$597235$ graphs.

\begin{table}[ht]
 \caption{For connected bipartite graphs with $v$ vertices and
     with minimum degree $3$ and maximum degree $4$, we have listed
 the number of graphs, the number of violations of the
 graph positivity, the average order $ng$ of the automorphism group of
 the graphs, the average order $ngv$ of this group for graphs
 violating the positivity property.}
\begin{tabular}{|c|c|c|c|c|}
 \hline
 $v$ & number of graphs & violations & $ng$ & $ngv$\\
 \hline
 12&    240&   1 & 14.1 & 192.\\
 14&    5183&   10&  4.3 & 143.\\
 16&    190378&  134 & 2.1 & 41.6\\
 18&    9816658&  6291 & 1.5  & 14.8\\
 \hline
 \end{tabular}
 \label{tabdeg34}
\end{table}

Let us now consider some non-regular bipartite lattice graphs.
In the case of rectangular grids of size $N_x \times N_y$,
we found few violations, occurring in the case of narrow grids.

In the case of rectangular grids with bc
 periodic only in the $y$ direction, we considered the cases
 in table \ref{tabny};

\begin{table}[ht]
 \caption{Sizes of the rectangular grids with bc periodic only in 
     the $y$ direction that have been tested. 
We have tested the positivity of the grids for all values of $N_x$ 
in the range indicated in the first row.}
\begin{tabular}{|c|c|c|c|c|c|c|c|c|c|}
\hline
$N_x$& 2-5600 & 2-2500& 2-1500 & 2-1000 & 2-300 & 2-120 & 2-50 & 2-23 \\ 
$N_y$& 4      & 6    &   8    &   10    & 12    & 14    & 16   & 18   \\ 
 \hline
 \end{tabular}
 \label{tabny}
\end{table}

The only positivity violations occur for $N_x \ge 421, N_y=4$
and $N_x=3, N_y=18$.

In the case of rectangular grids with open bc
we have considered the cases listed in the table \ref{tabsqnp}.
\begin{table}[ht]
    \caption{Tested rectangular grids $N_x\times N_y$ with open bc. }
\begin{tabular}{|c|c|}
\hline

$N_x \le 5000, N_y=2$  & $N_x \le 500, N_y=11$\\
$N_x \le 1800,  N_y=3$ & $N_x \le 500,  N_y=12$\\
$N_x \le 3700,  N_y=4$ & $N_x \le 180,  N_y=13$\\
$N_x \le 3800,  N_y=5$ & $N_x \le 100,  N_y=14$\\
$N_x \le 2700,  N_y=6$ & $N_x \le 70,  N_y=15$\\
$N_x \le 2700,  N_y=7$ & $N_x \le 45,  N_y=16$\\
$N_x \le 1700,  N_y=8$ & $N_x \le 50,  N_x=17$\\
$N_x \le 1400,  N_y=9$ & $N_x \le 35,  N_y=18$\\
$N_x \le 700,  N_y=10$ & $N_x \le 23, N_y=19$\\
 \hline
 \end{tabular}
 \label{tabsqnp}
\end{table}

We have found positivity violations only for $N_y=3$.

Summarizing, in the case of rectangular grids with any bc,
we have found no positivity violations for $N_y > 4$, $y$ being the shortest
direction.

In the case of cubic grids of sizes $N_x \times N_y \times N_z$
with open bc,  we have tested the positivity 
in the cases listed in the table \ref{tabcub}; there are no positivity
violations except in the case $N_x \ge 421, N_y=2, N_z=2$ which is isomorphic
to the rectangular grid $N_x \times 4$ periodic in the $y$-direction
considered in table \ref{tabny}.

\begin{table}[ht]
 \caption{Sizes of the cubic grids that have been tested for positivity.
We have tested the grids for all values of $N_x$ 
in the range indicated in the first row.}
\begin{tabular}{|c|c|c|c|c|c|c|c|c|c|}
\hline
$N_x$& 2-5600 & 3-3500& 4-2000 & 5-800 & 3-1600 & 4-500 & 5-150 & 4-85 & 5-13 \\ 
$N_y$& 2      & 3    &   4    &     5 & 3      & 4     & 5     & 4    & 5    \\ 
$N_z$& 2      & 2    &   2    &     2 & 3      & 3     & 3     & 4    & 4    \\
 \hline
 \end{tabular}
 \label{tabcub}
\end{table}

  As a final comment concerning
  biconnected non-bipartite graphs, it appears that the majority of these
  graphs do not have the graph-positivity property.
  We have checked that this is true for even $v$, $v \le 8$.
  For odd $v$, $v \le 9$ the frequency of violations increases rapidly and
  it is roughly $1/3$ for $v=9$.
  This is consistent with the fact that infinite non-bipartite lattices
  have been found to be not graph-positive[\onlinecite{bfp}].

  It would be interesting to investigate whether graphs which are not
  bipartite due to a few defects have the tendency to be positive.

  We have examined the positivity of a sequence of nanotubes,
  $C_{40+20N}$ in Appendix B for $1 \le N \le 300$.
  All these graphs are positive for $N > 7$.
  These graphs are formed by a nanotube composed of hexagons, and
  by two caps containing $6$ pentagons each, and some more hexagons.
  Roughly speaking, the nanotube's hexagon lattice tube tends to make the 
  graph positive,
  the pentagons negative, so if the nanotube is sufficiently long, 
  one can expect the graph to be positive.
  However it does not seem to be generally true that a graph formed
  by a subgraph which is positive and by a small non-bipartite part
  is positive. For instance consider a $(N_x, N_y)$ rectangular grid
  with open bc, with $N_y=5$. We saw that these
  graphs are positive (checked up to $N_y=3800$).
  Let us  create defects in the third and fourth vertical strips, by adding
  $SW-NE$ diagonals to its squares. We have considered these graphs up
  to $N_x=4000$: for $N_x \ge 10$, all these graphs violate positivity.
\section{Conclusions}
In a previous paper[\onlinecite{bfp}] we noticed that the dimer entropy of
infinite bipartite lattices seems to satisfy a positivity property.  We
have shown that this positivity property can be generalized to finite
graphs, as the positivity of the coefficients of the Newton expansion
for the "dimer entropy" of graphs.

We considered two natural classes of finite graphs related to infinite
bipartite lattices: the connected regular bipartite (RBB) graphs and finite
lattices with various bc.

In the first class, we have tested exhaustively a very large sample of 
RBB graphs.
The results of our survey lead us to conjecture that for all degrees $r$
in $r-$regular graphs, the frequency of the positivity violations tends to zero
as $v \to \infty$.  We have also observed that the graphs
showing positivity violations have an average order of the symmetry of
the automorphism group sizably larger than the average order for
graphs with the same number of vertices.

In the second class,
we have examined a large number of rectangular grids $N_x\times N_y$
with various bc.
We found no positivity violation for $min(N_x, N_y) > 4$.

In the three-dimensional case we could only test the positivity for
relatively small cubic grids (with open bc).  These results give
some support to the conjecture that for the infinite hypercubic
lattices in any dimension, with open bc, all the coefficients of the
series for the dimer entropy are positive[\onlinecite{bfp}].

In the case of hexagonal lattices, in the brick-wall representation
$N_x \times N_y$, we found no violations for $N_x \ge N_y$, except
for $N_x=N_y=4$, while there are several violations for $N_x < N_y$. 
 This may be related to the fact that the entropy of the
hexagonal lattice depends on the bc[\onlinecite{elser}].
Also in this case these computations suggest that when taking the
limit $N \to \infty$ for the $N \times N$ cases, the coefficients of the
Newton expansion for the dimer entropy of the infinite hexagonal
lattice are all positive.

{\bf Acknowledgments} We thank K. Markstr$\ddot{ \rm o}$m for discussions.
\section{Appendix A: Proof that $C_{2n}$ and $K_{v,v}$ 
are graph-positive}

A polygon $C_v$ with $v$ vertices has[\onlinecite{godsil}] 
$N(i) = \frac{v}{v-i}\binom{v-i}{i}$.
From (\ref{di}) one gets for $v$ even
\begin{equation}
d(i) = {\rm ln}\frac{(v-i-1)!(v-1)^i}{(v-1)!}
\end{equation}
from which
\begin{equation}
    \Delta d(i) = -{\rm ln}\left(1 - \frac{i}{v-1}\right) = 
\sum_{j=1}^{\infty} \frac{i^j}{j(v-1)^j}
\end{equation}
which is positive; and thus 
from $d(1) = 0$ one sees that $d(2), d(3), ...$  are positive.
\begin{equation}
    \Delta^{(k+1)} d(i) = \sum_{j=1} \frac{\Delta^{k}(i^j)}{j(v-1)^j}
\end{equation}
Observing that $i^j = \sum_{h=0}^j S_{j, h} (i)_h$, with $S_{j, h}$  the Stirling
numbers (of the second kind), and that $\Delta (i)_m = m(i)_{m-1}$, 
from the positivity of the Stirling numbers
and of the falling factorials, it follows that
$\Delta^{k} d(i)$ is positive for $k \ge 1$, so that $C_v$ is graph-positive
for $v$ even.

A complete bipartite graph $K_{n,n}$ has $N(i) = \binom{n}{i}^2i!$.

\begin{equation}
d(i) = {\rm ln}\left(\frac{n!}{(n-i)!(2n-1)(2n-3)...(2n-2i+1)}(\frac{2n-1}{n})^i\right)
\end{equation}
One has
\begin{equation}
d(i+1) = d(i) + {\rm ln}\frac{(n-i)(2n-1)}{(2n-2i-1)n} =
d(i) + {\rm ln}\left(1 + \frac{i}{(2n-2i-1)n}\right) > d(i)
\end{equation}
for $i < n$; since $d(1) = 0$ it follows that $d(i) \ge 0$ for $i < n$.
One has
\begin{equation}
\Delta d(i) = {\rm ln}\frac{(2n-1)(n-i)}{n(2n-2i-1)} =
{\rm ln}\left(1 + \frac{i}{n(2n-2i-1)}\right)
\end{equation}
so that $\Delta d(i) > 0$ for $i < n$.
From
\begin{equation}
    \Delta d(i) = {\rm ln}\frac{2n-1}{2n} - {\rm ln}\left(1 - \frac{1}{2(n-i)}\right) =
    {\rm ln}\frac{2n-1}{2n}  + \sum_{j=1}\frac{1}{j2^j}\frac{1}{(n-i)^j}
\end{equation}
it follows that for $m \ge 1$
\begin{equation}
\Delta^{m+1} d(i) = \sum_{j=1}\frac{1}{j2^j}\Delta^m\left(\frac{1}{(n-i)^j}\right)
\end{equation}
defined for $i \le n - m - 1$.
It is not difficult to show that these quantities are non-negative, observing that 
\begin{equation*}
 \Delta \prod_j\frac{1}{(n-i-n_j)}
\end{equation*}
where $0 \le n_j < k$, is the sum of terms of the form
\begin{equation*}
 \prod_j\frac{1}{(n-i-n'_j)}
\end{equation*}
where $0 \le n'_j < k+1$. Therefore $K_{n,n}$ is graph-positive.

Friedland, Krop and  Markstr$\ddot{ \rm o}$m[\onlinecite{FKM}] have computed an approximation for
the average value of $N(i)$ over all $0-1$ matrices of size $n\times n$ 
with row and column sums equal to $r$
\begin{equation}
    N^{Av}(i) = \frac{\binom{n}{i}^2 r^{2i} i!(r n - i)!}{(r n)!}
\end{equation}

One can show by the techniques of this section that the sequence of
$N^{Av}(i)$ leads to $d(i)$ satisfying graph positivity.
This is consistent with our conjecture that as $n \to \infty$
the frequency of violations of graph positivity goes to zero.
\section{Appendix B: A sequence of nanotubes}
A fullerene can be drawn on a sphere; it
has $12$ pentagonal faces, while the other faces are hexagons.
\begin{figure}[tbp]
\begin{center}
\leavevmode
\includegraphics[width=5.5 in]{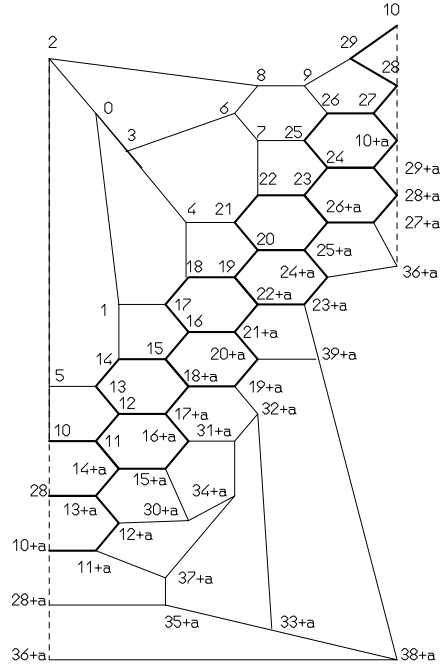}
\caption{ \label{figura2}  Nanotube with $N=1$; if the letter $a$
is present in a node index, add $20$ to the node index.}
\end{center}
\end{figure}

A nanotube $C_{40+20 N}$ can be formed by cutting a $C_{40}$ 
fullerene into a North cap and a South
cap, joined by a tube formed by hexagons, belonging to an hexagonal lattice,
formed by $N$ hexagonal strips with $20$ nodes
characterized by a "chiral vector" $(m,n)$ which describes the
periodicity conditions on the tube. In the case we have considered,
 $m=n=5$.(See Fig.1)

\begin{table}[ht]
    \caption{North cap}
\begin{tabular}{|c|c|c|c|c|c|c|c|}
 \hline
0-2& 2-8& 6-8& 3-6& 0-3& 0-1& 3-4& 6-7\\
8-9& 2-5& 4-18& 17-18& 1-17& 1-14& 16-17& 18-19\\
4-21& 5-13& 12-13& 13-14& 5-10& 20-21& 19-20& 21-22\\
7-22& 7-25& 22-23& 23-24& 24-25& 25-26& 9-26& 9-29\\
26-27& 10-29& 28-29& 27-28& 10-11& 11-12& 14-15& 15-16\\
 \hline
 \end{tabular}
 \label{tab6}
\end{table}
In Tables \ref{tab6}, \ref{tab7}, \ref{tab8} the edges must be read from
left to right, from up to down.  The edges are thus ordered in such a way
that the maximum number of 'active nodes'[\onlinecite{bpalg}] is $11$. 
\begin{table}[ht]
\caption{Strips: the first index in an edge $a-b$ stands for
$a + 20 i$ or for $a + 20(i+1)$ if it is followed by an $A$;
the second index $b$ stands for $a + 20(i+1)$;
$20$ is the number of nodes added by a strip;
$i=0,...,N-1$ where $N$ is the number of strips.
}
\begin{tabular}{|c|c|c|c|c|c|c|c|}
 \hline
11-14& 12-17&15-18&17A-18&16A-17\\
19-22&16-21&21A-22&20A-21&18A-19\\
19A-20&22A-23&20-25&23-26&25A-26\\
24A-25&23A-24&26A-27&24-29&27-10\\
10A-29&28A-29&27A-28&28-13&13A-14\\
10A-11&11A-12&12A-13&14A-15&15A-16\\
 \hline
 \end{tabular}
 \label{tab7}
\end{table}
\begin{table}[ht]
    \caption{South cap; to the indices of the edges one must add $20 N$}
\begin{tabular}{|c|c|c|c|c|}
 \hline
 11-37& 28-35& 35-37& 33-35& 34-37\\
 27-36& 24-36& 36-38& 33-38& 38-39\\
 23-39& 20-39& 32-33& 19-32& 31-32\\
 16-31& 31-34& 30-34& 12-30& 15-30\\
 \hline
\end{tabular}
\label{tab8}
\end{table}

In the algorithm[\onlinecite{bpalg}] for computing the matching
generating polynomial, the ordering of the edges is important for
performance.

We considered $1 \le N \le 300$;  graph positivity holds for $N > 7$.
\section{Appendix C: Examples of non-positive RBB graphs}
The non-positive RBB graph with $v=14$ and degree $3$ is the Heawood graph.
Its matching generating polynomial
coefficients $N(0),..., N(v/2)$ are 

$N = [1, 21, 168, 644, 1218, 1050, 336, 24]$.
The coefficients of $\bar{N}(0),..., \bar{N}(v/2)$ (Eq.\ref{nbar}) are

$\bar{N} = [1, 91, 3003, 45045, 315315, 945945, 945945, 135135]$.
From these and Eq.(\ref{di}) one computes $d(0), .., d(v/2)$.
The first negative finite difference is $\Delta^4 d(0) = -0.000558$.

To compute numerically the finite differences in Sage we used
$RealIntervalField(prec)$ with $prec=500$ for the RBB graphs, where
$prec$ is the number of bits of precision used;
for long rectangular grids higher precisions were required to have
enough significant digits.

There are two non-positive RBB graphs with $v=16$ and degree $3$;
the first one has reduced adjacency matrix

$[11100000, 11010000, 10001100, 01001100, 00001011, 00000111, 00110010, 00110001]$
and $N=[1, 24, 228, 1096, 2830, 3848, 2516, 632, 33]$;
the first negative finite difference is $\Delta^7 d(0) = -0.0029$.

The second one has reduced adjacency matrix

$[11100000, 10011000, 10000110, 01010001, 00110100, 01001010, 00001101, 00100011]$
and $N = [1, 24, 228, 1096, 2826, 3816, 2444, 600, 33]$;
The first negative finite difference is $\Delta^4 d(0) = -0.0000999$.

There are two non-positive RBB graphs with $v=22$ and degree $4$;
the first one has reduced adjacency matrix

$[11110000000, 11101000000, 11100100000, 11100010000, 00011101000, 00011010100,$

$00010110010, 00001110001, 00000001111, 00000001111, 00000001111]$

and $N = [1, 44, 814, 8272, 50675, 193516, 460870, 666512, 552265, 234972, 40968, 1584]$;
the first negative finite difference is $\Delta^9 d(0) = -0.00031$.

The second one has reduced adjacency matrix

$[11110000000, 11101000000, 11100100000, 11100010000, 00010001110, 00011100001,$

$00011010001, 00001110001, 00000101110, 00000011110, 00000001111]$

and $N=[1, 44, 814, 8272, 50678, 193576, 461295, 667872, 554256, 236160, 41076, 1584]$;
the first negative finite difference is $\Delta^9 d(0) = -0.000098$.

\end{document}